%% file: LoopInvs_ICCSA_2011.tex
\documentclass[10pt, conference, compsocconf]{IEEEtran}

\usepackage{etex}
\usepackage{delarray}
\usepackage{amssymb}
\usepackage{amsmath}
\usepackage{graphicx}
\usepackage{subfig}
\usepackage[usenames,dvipsnames]{color}
\usepackage[all]{xy}

\usepackage{aicescover}
\usepackage{url}

\input{rules}

\usepackage{float}
\floatstyle{boxed}
\newfloat{mybox}{htb}{lob}
\floatname{mybox}{Box}
\newsubfloat{mybox}

\usepackage{booktabs}
\usepackage{colortbl}

\usepackage{fancyvrb}

\usepackage{tikz}
\usetikzlibrary{shapes,snakes,arrows,shadows}

\definecolor{known}{RGB}{50, 205, 50} % green
\definecolor{unknown}{rgb}{1.0, 0.0, 0.0} % red
\definecolor{aicesred}{RGB}{152,26,37}
\definecolor{rwthblue}{RGB}{91,162,222}

\newcommand{\click}{{\sc{Cl\makebox[.58\width][c]{1}ck}}}
\newcommand{\lu}{$LU$ factorization}
\newcommand{\cs}{coupled Sylvester equation}
\newcommand{\prop}[2]{{\tt{#1(}}#2{\tt{)}}}

\usepackage{listings}
\usepackage{caption}
\DeclareCaptionFont{white}{\color{white}}
\DeclareCaptionFormat{listing}{\colorbox[cmyk]{0.43, 0.35, 0.35,0.01}{\parbox{0.47\textwidth}{\hspace{15pt}#1#2#3}}}
\newcommand{\LUGraphLevels}[1]
{
	\begin{tikzpicture}[scale=0.6,
					ball/.style={circle, fill=gray!70!white, draw=none, text=black, circular drop shadow},
					arrow/.style={->, shorten <= 1pt, >=stealth,thick}]
		\node[ball, fill=#1, at={( 0.0,  4.0)}] (n1) {1};
		\node[ball, fill=#1, at={(-2.0,  2.0)}] (n2) {2};
		\node[ball, fill=#1, at={( 2.0,  2.0)}] (n3) {3};
		\node[ball, fill=#1, at={( 0.0,  0.0)}] (n4) {4};
		\node[ball, fill=#1, at={( 0.0, -2.0)}] (n5) {5};
		\node[text=gray, at={(-4, 4)}]() {Level 1};
		\draw[-, thick, color=lightgray] (-4, 3) -- (4, 3);
		\node[text=gray, at={(-4, 2)}]() {Level 2};
		\draw[-, thick, color=lightgray] (-4, 1) -- (4, 1);
		\node[text=gray, at={(-4, 0)}]() {Level 3};
		\draw[-, thick, color=lightgray] (-4,-1) -- (4,-1);
		\node[text=gray, at={(-4, -2)}]() {Level 4};
		\draw[arrow] (n1.south) -- (n2.north);
		\draw[arrow] (n1.south) -- (n3.north);
		\draw[arrow] (n2.south) -- (n4.north);
		\draw[arrow] (n3.south) -- (n4.north);
		\draw[arrow] (n4.south) -- (n5.north);
	\end{tikzpicture}
}

\newcommand{\depGraphLU}[5]
{
	\begin{tikzpicture}[scale=0.6,
					ball/.style={circle, fill=gray!70!white, draw=none, text=black, circular drop shadow},
					arrow/.style={->, shorten <= 1pt, >=stealth,thick}]
		\node[ball, fill=#1, at={(-2.5,  0.9)}] (n1) {1};
		\node[ball, fill=#2, at={( 2.5,  0.9)}] (n2) {2};
		\node[ball, fill=#3, at={(-2.5, -0.9)}] (n3) {3};
		\node[ball, fill=#4, at={( 2.5, -0.9)}] (n4) {4};
		\node[ball, fill=#5, at={( 2.5, -2.6)}] (n5) {5};
		\draw[-, thick, color=lightgray] (-5,0) -- (5,0);
		\draw[-, thick, color=lightgray] (0,-2.0) -- (0,2.0);
		\draw[arrow] (n1.south) -- (n3.north);
		\draw[arrow] (n1.east)  -- (n2.west);
		\draw[arrow] (n3.east)  -- (n4.west);
		\draw[arrow] (n2.south) -- (n4.north);
		\draw[arrow] (n4.south) -- (n5.north);
	\end{tikzpicture}
}

\newcommand{\smallDepGraphLU}[5]
{
	\begin{tikzpicture}[%scale=0.6,
					ball/.style={circle, fill=gray!70!white, draw=none, text=black},%, circular drop shadow},
					arrow/.style={->, shorten <= 1pt, >=stealth,thick}]
		\node[ball, fill=#1, at={(-0.75,  0.32)}] (n1) {};
		\node[ball, fill=#2, at={( 0.75,  0.32)}] (n2) {};
		\node[ball, fill=#3, at={(-0.75, -0.32)}] (n3) {};
		\node[ball, fill=#4, at={( 0.75, -0.32)}] (n4) {};
		\node[ball, fill=#5, at={( 0.75, -0.88)}] (n5) {};
		\draw[-, thick, color=lightgray] (-1.5,0) -- (1.5,0);
		\draw[-, thick, color=lightgray] (0, -0.8) -- (0, 0.8);
		\draw[arrow] (n1.south) -- (n3.north);
		\draw[arrow] (n1.east)  -- (n2.west);
		\draw[arrow] (n3.east)  -- (n4.west);
		\draw[arrow] (n2.south) -- (n4.north);
		\draw[arrow] (n4.south) -- (n5.north);
	\end{tikzpicture}
}

\newcommand{\coupsylvDepGraph}[1]
{
	\begin{tikzpicture}[scale=0.6,
					ball/.style={circle, fill=gray!70!white, draw=none, text=black, circular drop shadow},
					arrow/.style={->, shorten <= 1pt, >=stealth,thick}]
		\node[ball, fill=#1, at={(-2.5,  4.3)}] (n1) {1};
		\node[ball, fill=#1, at={( 3.6,  2.6)}] (n2) {2};
		\node[ball, fill=#1, at={( 5.8,  2.6)}] (n3) {3};
		\node[ball, fill=#1, at={( 4.7,  0.9)}] (n4) {4};
		\node[ball, fill=#1, at={(-3.6, -0.9)}] (n5) {5};
		\node[ball, fill=#1, at={(-1.4, -0.9)}] (n6) {6};
		\node[ball, fill=#1, at={(-2.5, -2.6)}] (n7) {7};
		\node[ball, fill=#1, at={( 5.8, -4.6)}] (n8) {8};
		\node[ball, fill=#1, at={( 8.0, -4.6)}] (n10) {\footnotesize 10};
		\node[ball, fill=#1, at={( 1.4, -4.6)}] (n9) {9};
		\node[ball, fill=#1, at={( 3.6, -4.6)}] (n11) {\footnotesize 11};
		\node[ball, fill=#1, at={( 4.7, -6.3)}] (n12) {\footnotesize 12};
		\draw[-, thick, color=lightgray] (-5,0) -- (8,0);
		\draw[-, thick, color=lightgray] (0,-5.0) -- (0,5.0);
		\draw[arrow] (n1.east) -- (n2.north west);
		\draw[arrow] (n1.east) -- (n3.north west);
		\draw[arrow] (n2.south) -- (n4.north);
		\draw[arrow] (n3.south) -- (n4.north);
		\draw[arrow] (n1.south)  -- (n5.north);
		\draw[arrow] (n1.south)  -- (n6.north);
		\draw[arrow] (n4.south) -- (n8.north);
		\draw[arrow] (n4.south) -- (n10.north);
		\draw[arrow] (n5.south) -- (n7.north);
		\draw[arrow] (n6.south) -- (n7.north);
		\draw[arrow] (n7.east) -- (n9.north west);
		\draw[arrow] (n7.east) -- (n11.north west);
		\draw[arrow] (n8.south) -- (n12.north);
		\draw[arrow] (n9.south) -- (n12.north);
		\draw[arrow] (n10.south) -- (n12.north);
		\draw[arrow] (n11.south) -- (n12.north);
	\end{tikzpicture}
}

\newcommand{\smallDepGraphCoupSylv}[5]
{
	\begin{tikzpicture}[%scale=0.6,
					ball/.style={circle, fill=gray!70!white, draw=none, text=black},% circular drop shadow},
					arrow/.style={->, shorten <= 1pt, >=stealth,thick}]
		\node[ball, fill=aicesred,  at={(-0.40,  1.20)}] (n1)  {};
		\node[ball, fill=#3, at={(-0.60, -0.20)}] (n5)  {};
		\node[ball, fill=#4, at={(-0.20, -0.20)}] (n6)  {};
		\node[ball, fill=#5, at={(-0.40, -0.70)}] (n7)  {};
		\node[ball, fill=#1, at={( 0.60,  0.70)}] (n2)  {};
		\node[ball, fill=#2, at={( 1.00,  0.70)}] (n3)  {};
		\node[ball, fill=#5, at={( 0.80,  0.20)}] (n4)  {};
		\node[ball, fill=#5, at={( 0.20, -1.20)}] (n9)  {};
		\node[ball, fill=#5, at={( 0.60, -1.20)}] (n11) {};
		\node[ball, fill=#5, at={( 1.00, -1.20)}] (n8)  {};
		\node[ball, fill=#5, at={( 1.40, -1.20)}] (n10) {};
		\node[ball, fill=rwthblue, at={( 0.80, -1.70)}] (n12) {};
		\draw[-, thick, color=lightgray] (-1.0,0) -- (1.6,0);
		\draw[-, thick, color=lightgray] (0,-2.0) -- (0,1.5);
		\draw[arrow] (n1.east) -- (n2.north west);
		\draw[arrow] (n1.east) -- (n3.north west);
		\draw[arrow] (n2.south) -- (n4.north);
		\draw[arrow] (n3.south) -- (n4.north);
		\draw[arrow] (n1.south)  -- (n5.north);
		\draw[arrow] (n1.south)  -- (n6.north);
		\draw[arrow] (n4.south) -- (n8.north);
		\draw[arrow] (n4.south) -- (n10.north);
		\draw[arrow] (n5.south) -- (n7.north);
		\draw[arrow] (n6.south) -- (n7.north);
		\draw[arrow] (n7.east) -- (n9.north west);
		\draw[arrow] (n7.east) -- (n11.north west);
		\draw[arrow] (n8.south) -- (n12.north);
		\draw[arrow] (n9.south) -- (n12.north);
		\draw[arrow] (n10.south) -- (n12.north);
		\draw[arrow] (n11.south) -- (n12.north);
	\end{tikzpicture}
}

\begin{document}

\title{Automatic Generation of Loop-Invariants\\for Matrix Operations}
\author{\IEEEauthorblockN{Diego Fabregat-Traver and Paolo Bientinesi}
\IEEEauthorblockA{
AICES, RWTH Aachen\\
Aachen, Germany\\
\{fabregat,pauldj\}@aices.rwth-aachen.de}
}

\aicescoverauthor{Diego Fabregat-Traver and Paolo Bientinesi}
\aicescoverpublisher{\normalsize{\bf * The final version published by IEEE is available at:}\\
                    \url{http://www.computer.org/csdl/proceedings/iccsa/2011/4404/00/4404a082-abs.html}}
\aicescoverpage

\maketitle

\begin{abstract}
In recent years it has been shown that for many linear algebra
operations it is possible to create families of algorithms following a
very systematic procedure. We do not refer to the fine tuning of a
known algorithm, but to a methodology for the actual generation of
both algorithms and routines to solve a given target matrix
equation. Although systematic, the methodology relies on complex
algebraic manipulations and non-obvious pattern matching, making the
procedure challenging to be performed by hand; our goal is the
development of a fully automated system that from the sole description
of a target equation creates multiple algorithms and routines. We
present \click{}, a symbolic system written in Mathematica, that
starts with an equation, decomposes it into multiple equations, and
returns a set of loop-invariants for the algorithms---yet to be
generated---that will solve the equation. In a successive step each
loop-invariant is then mapped to its corresponding algorithm and
routine. For a large class of equations, the methodology
generates known algorithms as well as many previously
unknown ones. Most interestingly, the methodology unifies
algorithms traditionally developed in isolation. As an example,
the five well known algorithms for the \lu{} are for 
the first time unified under a common root.
\end{abstract}


\section{Introduction}

In order to attain high-performance on a variety of architectures and
programming paradigms, for a target operation not one but multiple
algorithms are needed. We focus our attention on the domain of matrix
equations and aim for a symbolic system, fully automated, that takes
as input the description of an equation $Eq$ and returns algorithms and
routines to solve $Eq$.

\begin{mybox}
  \scriptsize
  \centering
  \renewcommand{\arraystretch}{1.6}
  $$
\left(
          \begin{array}{@{\,}c@{\,}}
            \{X_{T}, Y_{T} \} = \Psi(A_{TL}, B, C_{T}, D_{TL}, E, F_{T}) \\\hline
            \{X_{B}, Y_{B} \} = \Psi(A_{BR}, B, C_{B} - A_{BL} X_{T}, D_{BR}, E, F_{B} - D_{BL} X_{T})
          \end{array}
\right)
  $$ \\
  \caption{Partitioned Matrix Expression for the \cs{}.} \label{box:PMEEx}
\end{mybox}

\begin{figure}
  \begin{center}
    \includegraphics[scale=0.75]{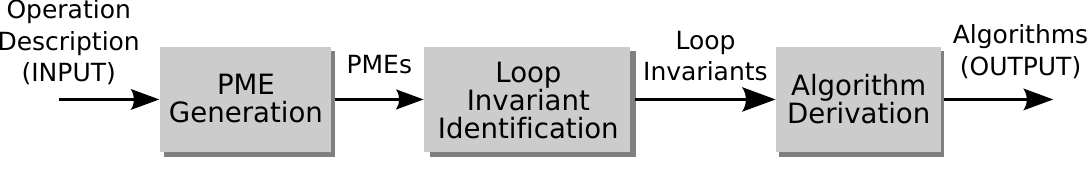}
    \caption{The process of algorithm generation can be broken down
      into three main stages.} \label{fig:steps}
  \end{center}
\end{figure}

This research is inspired by an existing methodology for the
derivation of families of algorithms, which is based on formal methods
and program correctness ~\cite{Bientinesi:2005:SDD,PaulDj:PhD-TR}.  As
depicted in Fig.~\ref{fig:steps}, in the process of algorithm
generation we identify three successive stages: ``PME Generation'',
``Loop-Invariant Identification'', and ``Algorithm Derivation''.  The
input to the process is the description of a target operation. In the
first stage, the {\em Partitioned Matrix Expression} (PME) for the
operation is obtained. 
A PME is a decomposition of the original problem into simpler sub-problems
in a ``divide and conquer'' fashion, exposing the computation
to be performed in each part of the output matrices.
As an example, in Box~\ref{box:PMEEx}
we show the PME for the coupled Sylvester equation:

{
\footnotesize
$$
\begin{array}{c@{\quad\!}c@{\quad\!}c}
\begin{aligned}
\{X, Y\} = \Psi(&A,B,C,\\&D,E,F)
\end{aligned} &
\equiv &
\left\{ 
  \begin{array}{@{}l@{}}
    A X + Y B = C \\
    D X + Y E = F 
  \end{array} 
\right.
\end{array}
$$
}

The second stage of the process deals with the identification of
boolean predicates, the {\em Loop-Invariants}~\cite{GrSc:92}, that
describe the intermediate state of computation for the sought-after
algorithms. 
Loop-invariants can be extracted from the PME, and are at
the heart of the automation of the third stage.
Box~\ref{box:LoopInvEx} contains an example of
loop-invariant.
\begin{mybox}
  \centering
  \small
  \renewcommand{\arraystretch}{1.4}
  $$
\left(
          \begin{array}{@{\,}c@{\,}}
            \{X_{T}, Y_{T} \} = \Psi(A_{TL}, B, C_{T}, D_{TL}, E, F_{T}) \\\hline
            \neq
          \end{array}
\right)
  $$ \\
  \caption{One of the loop-invariants for the \cs{}. 
           The symbol $\neq$ 
           indicates that no constraints on the contents of the 
		   variables are imposed.
		   } \label{box:LoopInvEx}
\end{mybox}

In the third and last stage of the methodology, each loop-invariant is
transformed into its corresponding loop-based algorithm.  This stage
makes use of classical concepts in computer science such as formal
program correctness, Hoare's triples, and the invariance theorem.

\begin{figure}
  \centering
  \includegraphics[scale=0.75]{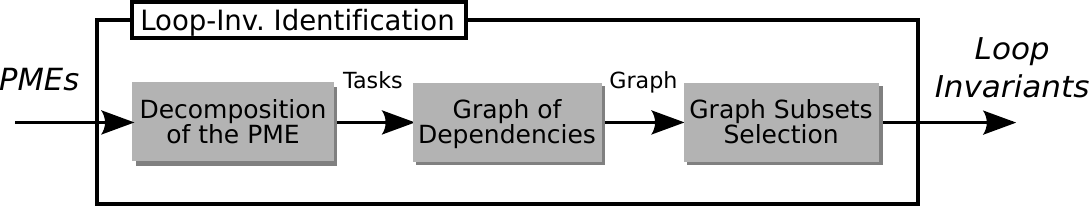}
  \caption{Steps for the identification of loop-invariants from a PME.} \label{fig:stepsLInv}
\end{figure}

We consider this paper as the second in a series. In the first
one~\cite{CASC-2011-PME} we introduced \click{}, a symbolic system written in
Mathematica~\cite{MathematicaOnline}, for the automatic generation of algorithms. There we
detailed how \click{} makes use of rewrite rules and pattern matching
to automatically generate PMEs from the description of target operations.
This paper centers around the second stage of the derivation process, the
Loop-Invariant Identification.
We describe the necessary steps to obtain a family of loop-invariants
from a given PME, Fig.~\ref{fig:stepsLInv}, and expose how \click{}
automates them through an extensive usage of pattern matching and
rewrite rules.

As in the example in Box~\ref{box:PMEEx}, a PME decomposes the target
operation into a set of equalities. Each of the equalities expresses
the computation to be carried out in the different parts of the matrix
to compute the overall equation.
Since an equality may represent a complex operation, we first
decompose it into a sequence of tasks.
We define tasks as basic units of computation matched by simple
patterns such as $C = A + B$, $C = A B$, $B = A^{-1}$ or $X = A^{-1} B$.
Next, we inspect the tasks for dependencies among them, and
build the corresponding dependency graph. Then, 
predicates that are candidates to becoming loop-invariants are identified
as subsets of the graph satisfying the dependencies.  Such subgraphs
represent tasks included in the equalities and, therefore, are
equivalent to choosing subsets of the computation included in the PME.
In the final step, the candidate predicates are checked for
feasibility and the resulting ones are labelled as viable loop-invariants.

The methodology described in~\cite{PaulDj:PhD-TR} generates loop-based algorithms 
that all share a fixed structure: 
a basic initialization followed by a loop in which the actual computation
is carried out (Box~\ref{box:skeleton}). The main idea of the
methodology is to identify a loop-invariant on top of which a proof
of correctness is built. Quoting Gries and Schneider from their
book {\em A Logical Approach to Discrete Math}~\cite{GrSc:92}

\vspace{2mm}
{\em ``Loop-invariants are crucial to understanding loops---so crucial that all
but the most trivial loops should be documented with the invariants used
to prove their annotations correct. In fact, (a first approximation to) the
invariant should be developed \textbf{before} the loop is written and should act as
a guide to the development of the loop.''}
\vspace{2mm}

A loop-invariant has to be satisfied
before the loop is entered and at the top and the bottom of each
iteration. Upon completion of the loop, the loop-invariant as
well as the negation of the loop-guard are satisfied. 
Given these known facts, the statements of the algorithms are
chosen to satisfy them. In particular, the loop-invariant, $LI$,
and the loop-guard, $G$, must be chosen so that $LI \wedge \neg G$
implies that the target equation has been solved.

\input{inputs/skeleton}

As the complexity of the target equation increases,
the methodology requires longer and more involved
algebraic manipulation and pattern matching,
making the manual generation of algorithms a tedious and
error-prone process. The situation is aggravated by
the fact that not one but multiple algorithms
are desired for one same target equation. For this
reason we advocate an automated symbolic system
which exploits the capabilities of modern computer
algebra tools to carry out the entire derivation
process.

In this paper, we make progress towards such a vision
detailing how \click{} performs all the steps involved in the 
Loop-Invariant Identification. The paper is organized as follows. 
In Section~\ref{sec:input} we illustrate the formalism used to
describe the target operations. The automatic generation
of PMEs is reviewed in Section~\ref{sec:PME}.
In Section~\ref{sec:linv} we detail how loop-invariants are
identified and how the process is automated, while
in section~\ref{sec:coupsylv} a more challenging example
is treated. We draw conclusions in Section~\ref{sec:conclusions}.

\section{Input to Cl1ck} \label{sec:input}

In line with the methodology we follow for the derivation
of algorithms, we choose the formalism 
traditionally used to reason about program correctness: 
operations shall be specified by means of the predicates Precondition ($P_{\rm pre}$) and
Postcondition ($P_{\rm post}$)~\cite{GrSc:92}. The precondition enumerates
the operands that appear in the equation and describes their properties, while
the postcondition specifies the equation that combines the operands.

As an example, Box~\ref{box:LUOpDesc} contains the description of the \lu{}.
The precondition states that the unit-diagonal, lower triangular matrix $L$ 
and the upper triangular matrix $U$ are unknown, and $A$ is an input matrix
for which the \lu{} exists. The postcondition indicates that, when the computation
completes, the product $L U$ equals $A$; while 
the notation $\{L, U\} = LU(A)$ denotes that $L$ and $U$ are the $LU$ factors of $A$.

\begin{mybox}
\footnotesize
$$
\{L, U\} = LU(A) \equiv
\left\{
\begin{split}
P_{\rm pre}: \{ & \prop{Unknown}{L} \, \wedge \, \prop{LowTri}{L} \, \wedge \\
				& \prop{UnitDiag}{L} \, \wedge \\
				& \prop{Unknown}{U} \, \wedge \, \prop{UppTri}{U} \, \wedge \\
                & \prop{Known}{A} \, \wedge \, \prop{\exists \; LU}{A} \} \\
\\
P_{\rm post}: \{ &  L U = A \}
\end{split}
\right.
$$
\caption{Formal description for the \lu{}.}
\label{box:LUOpDesc}
\end{mybox}

The two predicates in Box~\ref{box:LUOpDesc} describe unambiguously
the \lu{} and characterize the only knowledge about the operation
needed by \click{} to automate the generation of algorithms.
Box~\ref{box:MathDesc} illustrates the corresponding Mathematica 
statements required from the user.

\begin{mybox}
\footnotesize
\begin{Verbatim}[commandchars=+\$\%]

+textbf$precondition% = {
    { L, {"Output", "Matrix", "LowerTriangular",
          "UnitDiagonal"} },
    { U, {"Output", "Matrix", "UpperTriangular"} },
    { A, {"Input",  "Matrix", "ExistsLU"} }
};
+textbf$postcondition% = {
    { equal[times[L , U], A] } (* L U = A *)
};
\end{Verbatim}
\caption{Mathematica representation of the precondition and postcondition predicates for the \lu{}.}
\label{box:MathDesc}
\end{mybox}

We use the pair of predicates, $P_{\rm pre}$ and $P_{\rm post}$, to describe
every target operation. Such a description is the input to the generation of PMEs
and, therefore, to the whole process of algorithms derivation.

\section{Generation of PMEs} \label{sec:PME}

Having established a formalism to input a target operation, here we
summarize the process of PME generation. Since the objective is a
{\em Partitioned} Matrix Expression, \click{} starts off by rewriting
the equation in the postcondition in terms of partitioned matrices. To
this end, we introduce a set of rules to partition operands.  As shown
in Box~\ref{box:part}, a generic matrix $A$ can be partitioned in four
different ways. 
For a vector, only the $2 \times 1$ and $1 \times 1$ rules apply,
while for scalars only the $1 \times 1$ rule is admissible.

\begin{mybox}
\small
\renewcommand{\arraystretch}{1.4}
\vspace{2mm}
\begin{center}
    \subfloat[$2 \times 2$ rule]{
        \label{sbox:part2x2}
        \begin{minipage}{3.6cm}
		  \centering
		  $\ruleTwoByTwo{A}{m}{n}{TL}{k_1}{k_2}$
        \end{minipage}
    }
    \qquad
    \subfloat[$2 \times 1$ rule]
    {\label{sbox:part2x1}
        \begin{minipage}{3.6cm}
		  \centering
		  $\ruleTwoByOne{A}{m}{n}{T}{k_1}{n}$
        \end{minipage}
    }
    \\
    \subfloat[$1 \times 2$ rule]
    {\label{sbox:part1x2}
        \begin{minipage}{3.6cm}
		  \centering
		  $\ruleOneByTwo{A}{m}{n}{L}{m}{k_2}$
        \end{minipage}
    }
    \qquad
    \subfloat[$1 \times 1$ rule]
    %\subfloat[$1 \times 1$ (identity) rule]
    {\label{sbox:part1x1}
        \begin{minipage}{3.6cm}
		  \centering
		  $\ruleOneByOne{A}{m}{n}$
        \end{minipage}
    }

\end{center}
 
\caption{
  Rules for  partitioning  a generic matrix operand A.
  We use the subscript letters $T$, $B$, $L$, and $R$ for $T$op, $B$ottom,
  $L$eft, and $R$ight, respectively.
}\label{box:part}
\end{mybox}

The partitionings for an operand are constrained not only by its type
(matrix, vector or scalar) but also by its structure: if the operand presents a known structure,
such as triangularity or symmetry, we restrict the viable
partitionings to those that allow the inheritance of properties.  For
instance, Box~\ref{box:partL} illustrates the admissible partitionings
for a lower triangular matrix $L$. Only two rules allow the
inheritance: when the $1 \times 1$ rule is applied, $L$ remains unchanged, and
therefore triangular; a constrained $2 \times 2$ rule in which the
$TL$ quadrant is square leads a partitioning where both $L_{TL}$ and
$L_{BR}$ are square and lower triangular, $L_{TR}$ is zero, and
$L_{BL}$ is a generic matrix.

\begin{mybox}
\small
\renewcommand{\arraystretch}{1.4}
\vspace{2mm}
\begin{center}
         \begin{tabular}{lcl}
        \begin{minipage}{3.6cm}
		  \centering
		  $\lowtriRuleTwoByTwo{L}{m}{m}{TL}{k}{k}$
        \end{minipage}
         \quad or \qquad &
        \begin{minipage}{3.3cm}
		  \centering
		  $\ruleOneByOne{L}{m}{m}$
        \end{minipage}
         \end{tabular}
%}
\caption{Partitioning rules for a lower triangular matrix $L$.}\label{box:partL}
\end{center}
\end{mybox}

Finally, the viable partitionings are also constrained by the
operators that appear in the postcondition. For instance, in the
\lu{}, the operator {\em times} in $L U$ imposes that if $L$ is
partitioned along the columns, then $U$ has to be partitioned along
the rows and vice versa, so that the product is well defined.  Since
the set of rules where all the operands are partitioned $1 \times 1$
does not lead to a {\em Partitioned} Matrix Expression, the only
admissible set of partitioning rules for the \lu{} is shown in Box~\ref{box:LUPart}.
An efficient algorithm that identifies all the admissible partitioning rules
for a given equation was introduced in~\cite{CASC-2011-PME}.

\begin{mybox}
\small
\renewcommand{\arraystretch}{1.4}
$$
\begin{array}{c}
	\lowtriRuleTwoByTwo{L}{m}{m}{TL}{k}{k} \raisebox{2.8mm}{\textnormal{, }} \hspace{5mm}
	\upptriRuleTwoByTwo{U}{m}{m}{TL}{k}{k} \raisebox{2.4mm}{\textnormal{ and }} \\[10mm]
	\ruleTwoByTwo{A}{m}{m}{TL}{k}{k}
\end{array}
$$
\caption{Set of partitioning rules for the \lu{}.} \label{box:LUPart}
\end{mybox}

Once the valid partitioning rules are found, \click{} applies them to
the postcondition to obtain a predicate called {\em
partitioned postcondition}. In the case of the 
\lu{},
the corresponding partitioned postcondition is

{\small
\vspace{-2mm}
$$
\begin{array}{l}
  \label{eqn:matArit1}
  L U = A
  \; \Rightarrow % \\[2mm]
  \;\renewcommand{\arraystretch}{1.2}
  \left( \begin{array}{@{}c@{\,}|@{\,}c@{}} L_{TL} & 0 \\\hline L_{BL} & L_{BR} \end{array} \right)
  \left( \begin{array}{@{}c@{\,}|@{\,}c@{}} U_{TL} & U_{TR} \\\hline 0 & U_{BR} \end{array} \right)
  =
  \left( \begin{array}{@{}c@{\,}|@{\,}c@{}} A_{TL} & A_{TR} \\\hline A_{BL} & A_{BR} \end{array} \right)
\end{array}.
$$
}

\noindent
From here, matrix arithmetic is carried out until 
the equality operator is distributed over the partitions
yielding a set of equalities, one per quadrant:

{\footnotesize
\vspace{-1mm}
\begin{eqnarray}
  \label{eqn:matArit1}
  \renewcommand{\arraystretch}{1.4}
  \left( \begin{array}{@{}c@{\,}|@{\,}c@{}} L_{TL} & 0 \\\hline L_{BL} & L_{BR} \end{array} \right)
  \left( \begin{array}{@{}c@{\,}|@{\,}c@{}} U_{TL} & U_{TR} \\\hline 0 & U_{BR} \end{array} \right)
  =
  \left( \begin{array}{@{}c@{\,}|@{\,}c@{}} A_{TL} & A_{TR} \\\hline A_{BL} & A_{BR} \end{array} \right)
  \quad \Rightarrow %\quad
  \notag \\
  \renewcommand{\arraystretch}{1.4}
  \left( \begin{array}{@{}c@{\,}|@{\,}c@{}} 
  	L_{TL} U_{TL} & 
    L_{TL} U_{TR} \\\hline 
	L_{BL} U_{TL} &
	L_{BL} U_{TR} + L_{BR} U_{BR}
  \end{array} \right)
  =
  \left( \begin{array}{@{}c@{\,}|@{\,}c@{}} A_{TL} & A_{TR} \\\hline A_{BL} & A_{BR} \end{array} \right)
  \quad \Rightarrow %\quad
  \notag \\
  \renewcommand{\arraystretch}{1.4}
  \left( \begin{array}{@{}c@{\,}|@{\,}c@{}} 
    L_{TL} U_{TL} = A_{TL} &
    L_{TL} U_{TR} = A_{TR} \\\hline
    L_{BL} U_{TL} = A_{BL} &
    L_{BL} U_{TR} + L_{BR} U_{BR} = A_{BR}
  \end{array} \right) .  
\end{eqnarray}
}

At this point, an iterative process involving algebraic manipulation
and pattern matching transforms Eq.~\ref{eqn:matArit1} into the 
sought-after PME.  Central to this step is the capability of \click{} to
learn the pattern that defines the target operation.  Initially
\click{} only knows the pattern for a set of basic operations:
addition, multiplication, inversion and transposition.  This
information is hard-coded. More patterns are discovered while tackling
new operations.  For instance, the definition of the \lu{} in
Box~\ref{box:LUOpDesc} defines a pattern.  The pattern establishes
that two matrices $X$ and $Y$ are the $LU$ factors of a matrix $Z$ if
the constraints in the precondition are satisfied, and $X$, $Y$ and
$Z$ are related as dictated by the postcondition ($X Y = Z$).  Such
patterns provide \click{} with the necessary knowledge to identify
known operations within each of the equalities in
Eq.~\ref{eqn:matArit1}. 

Thanks to the inheritance of properties, the system recognizes that
the matrices $L_{TL}$, $U_{TL}$ and $A_{TL}$ match the pattern in
Box~\ref{box:LUOpDesc}, and therefore asserts that $ \{ L_{TL}, U_{TL}
\} = LU(A_{TL}) $. Similarly, \click{} identifies that $U_{TR}$ and
$L_{BL}$ result from two triangular systems, and that $L_{BR}$ and
$U_{BR}$ are the $LU$ factors of an updated matrix $A_{BR}$.
Box~\ref{box:PMELU} contains the outcome of this process, the PME for the \lu{}.
Notice that no restrictions on the size of the sub-operands was imposed;
the decomposition expressed by the PME is valid independently of the size of
the sub-operands, provided that $L_{TL}$, $U_{TL}$ and $A_{TL}$ are square.

\begin{mybox}
  \scriptsize
  \centering
  \renewcommand{\arraystretch}{1.6}
  $$
  \left( {\begin{array}{@{\;}c@{\;}|@{\;}c@{\;}} 
	  \{ L_{TL}, U_{TL} \} = LU(A_{TL})   &    U_{TR} = L_{TL}^{-1} A_{TR} \\\hline 
	     L_{BL} = A_{BL} U_{TL}^{-1}      & \{ L_{BR}, U_{BR} \} = LU(A_{BR} - L_{BL} U_{TR}) 
  \end{array}} \right)
  $$ \\
  \caption{Partitioned Matrix Expression for the \lu{}.} \label{box:PMELU}
\end{mybox}

\section{Identification of Loop-Invariants} \label{sec:linv}

Loop-invariants are the key predicates to prove the correctness of
loop-based algorithms. A loop-invariant expresses the state of the
variables as the computation unfolds.  Since a PME encapsulates the
computation to be performed to solve a target equation, our approach
identifies loop-invariants as subsets of the operations included in
the PME.

The Loop-Invariant Identification process consists
on three steps:
1) \click{} inspects each of the equalities
included in the PME and decomposes them into a sequence of tasks,
i.e., basic units of computation;
2) an analysis of the tasks yields the
dependencies among them, leading to a graph of dependencies where the
nodes are the tasks and the edges are the dependencies;
3) \click{} traverses the graph selecting all possible subgraphs satisfying the
dependencies. The subgraphs correspond to predicates that are candidates to
becoming loop-invariants. \click{} checks the feasibility of such predicates,
discarding the non-feasible ones and promoting the remaining ones to loop-invariants.

\subsection{Decomposition of the PME}

\click{} commences by analyzing the equalities in the PME.
Each equality satisfies a canonical form where the left-hand
side contains the output sub-operand(s) and the right-hand side
the explicit computation to obtain the output quantity(ies).
The right-hand side may be expressed either as a combination 
of sub-operands and the basic operators (plus, times, transpose,
inverse) or as an explicit function with one
or more input arguments. In this first step \click{}
decomposes the right-hand side of each equality into a 
sequence of one or more tasks.

The decomposition is led by a set of rules based on
pattern matching to identify whether an expression
is a basic task or a complex computation. In the case
of a complex computation, such rules also express how
to decompose it into simpler expressions. In this and
following sections we use the examples to illustrate the
decomposition rules.

We start the discussion with the \lu{} example. 
As Box~\ref{box:PMELU} shows, its PME comprises 
four equalities. 
The decomposition of equalities can be performed
independently from one another; \click{} arbitrarily
traverses the equalities by rows.  
The analysis commences from the top-left quadrant:
{\small $\{L_{TL}, U_{TL}\} = LU(A_{TL}).$}
Since the right-hand side matches a pattern 
associated to a basic task
{\small $\tt{f(x_0, x_1, \ldots, x_n)} \wedge
\forall{x_i}\, | \tt{isSuboperandQ}(x_i)$}, 
a function where all
the input arguments are (sub-)operands,
no decomposition
is necessary and the system only returns one task:
{\small $\{L_{TL}, U_{TL}\} := LU(A_{TL}).$}

The analysis procedes with the top-right quadrant:
{\small $U_{TR} = L_{TL}^{-1} A_{TR}.$}
The expression is matched by 
the pattern {\small $\tt{X = A^{-1} B} \wedge 
\tt{isLowTriQ}(A) \wedge \tt{NonSingularQ}(A)$}
and corresponds to the solution of a triangular system of equations.
\click{} recognizes it as a basic task and returns it.
Similarly for the bottom-left quadrant in which a third task
is identified.

Only one equality remains to be studied:
{\small $\{L_{BR}, U_{BR}\} = LU(A_{BR} - L_{BL} U_{TR})$}. The
expression matches the pattern
{\small $\tt{f(x_0, x_1, \ldots, x_n) \wedge
\exists{x_i} | \neg{\tt{isSuboperandQ}}(x_i)}$},
meaning that at least one of the input arguments is not
a (sub-)operand. Each complex argument is, therefore,
recursively analyzed to identify a sequence of basic tasks.
In the example,
{\small $A_{BR} - L_{BL} U_{TR}$} is the only complex argument;
it is matched by the pattern {\small $\tt{A - BC}$}, corresponding
to a basic task. As a result, \click{} yields the list
\{{\small $A_{BR} := A_{BR} - L_{BL} U_{TR}$,
$\{L_{BR}, U_{BR}\} := LU(A_{BR})$}\}.
In total, the algorithm produces the following five tasks:

\begin{enumerate}
\small
\item $\{L_{TL}, U_{TL}\} := LU(A_{TL});$ \\[-2mm]
\item $U_{TR} := L_{TL}^{-1} A_{TR};$ \\[-2mm]
\item $L_{BL} := A_{BL} U_{TL}^{-1};$ \\[-2mm]
\item $A_{BR} := A_{BR} - L_{BL} U_{TR};$ \\[-2mm]
\item $\{L_{BR}, U_{BR}\} := LU(A_{BR}).$ \\[-2mm]
\end{enumerate}

\subsection{Graph of dependencies}

Once the decomposition into tasks is available, \click{} proceeds with the study of
the dependencies among them. Three different kinds of dependencies may occur.

\begin{itemize}
\item {\bf True dependency.} One of the input arguments of a task is 
also the result of a previous task:
$$
  \begin{array}{@{}c@{\;}c@{\;}c@{}}
    A & := & B + C \\
    X & := & A + D
  \end{array}
$$
The order of the updates cannot be reversed because the second
one requires the value of $A$ computed in the first one.

\item {\bf Anti dependency.} One of the input arguments of a task is 
also the result of a subsequent task:
$$
  \begin{array}{@{}c@{\;}c@{\;}c@{}}
    X := A + D \\
    A := B + C 
  \end{array}
$$
The order of the updates cannot be reversed because the
first update needs the value of $A$ before the second one overwrites it.

\item {\bf Output dependency.} The result of a task is 
also the result of a different task:
$$
  \begin{array}{@{}c@{\;}c@{\;}c@{}}
    A := B + C \\
    A := D + E
  \end{array}
$$
The second update cannot be performed until the first is computed to
ensure the correct final value of $A$.
\end{itemize}

At a first sight, in the context of PMEs, 
it is difficult to distinguish between true and
anti dependencies since there is no clear order
in the execution. However, since each equality refers to 
the computation of a different part of the output
matrices, any time the output of an equality is
found as an input argument of another one, it
implies a true dependency: first the quantity is computed,
then it is used in a different equality.

Also, for the same reason, it is not easy to
distinguish the direction of an output dependency.
Since output dependencies only occur among
tasks belonging to the same equality (each
equality writes to a different part of the output
matrices), the order is determined because one of 
the involved tasks comes from the decomposition of the other
one, imposing an order in their execution.
While in general all three types of dependencies may appear, 
in the examples we provide only true dependencies arise.

We detail the analysis of the dependencies following the
example of the \lu{}.
During the analysis we use {\bf boldface} to highlight
the dependencies. 
The study commences with Task 1, 
whose output is {\small $\{L_{TL}, U_{TL}\}$}. \click{}
finds that the sub-operands $L_{TL}$ and $U_{TL}$
are input arguments for 
Tasks 2 and 3, respectively.

\begin{enumerate}
\item $\mathbf{\{L_{TL}, U_{TL}\}} := LU(A_{TL})$ \\[-3mm]
\item $U_{TR} := \mathbf{L_{TL}}^{-1} A_{TR}$ \\[-3mm]
\item $L_{BL} := A_{BL} \mathbf{U_{TL}}^{-1}$ \\[-3mm]
\end{enumerate}
This means that two true dependencies exist: one from 
Task 1 to Task 2 and another from Task 1 to Task 3.
Next, \click{} inspects Task 2, whose output is
$U_{TR}$. $U_{TR}$ is also identified as input 
for Task 4. 
\begin{enumerate}
\setcounter{enumi}{1}
\item $\mathbf{U_{TR}} := L_{TL}^{-1} A_{TR}$ \\[-2mm]
\setcounter{enumi}{3}
\item $A_{BR} := A_{BR} - L_{BL} \mathbf{U_{TR}} $ \\[-3mm]
\end{enumerate}
Hence, a true dependency from Tasks 2 to 4 is
imposed.
A similar situation arises when inspecting
Task 3, originating a true dependency from Task 3 to Task 4.

The analysis continues with Task 4; this computes an update
of $A_{BR}$, which is then used as input by Task 5,
thus, creating one more true dependency.
\begin{enumerate}
\setcounter{enumi}{3}
\item $\mathbf{A_{BR}} := A_{BR} - L_{BL} U_{TR} $ \\[-2mm]
\item $\{L_{BR}, U_{BR}\} := LU(\mathbf{A_{BR}}) $ \\[-2mm]
\end{enumerate}
Task 5 remains to be analyzed. Since its output, $\{L_{BR}, U_{BR}\}$, 
does not appear in any of the other tasks, no new dependencies are
found.

\begin{figure}
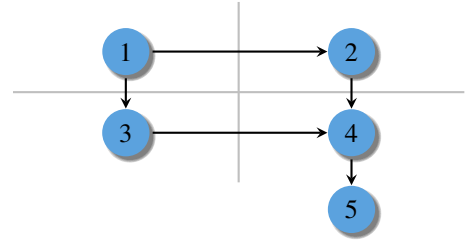

	\centering
	\depGraphLU{rwthblue}{rwthblue}{rwthblue}{rwthblue}{rwthblue}
	\caption{Final graph of dependencies for the \lu{}.} \label{fig:luGraph}
\end{figure}

In Fig.~\ref{fig:luGraph}, the list of the dependencies for the \lu{}
are mapped onto the graph in which 
node $i$ represents Task $i$.

\subsection{DAG subsets selection} \label{sec:depGraph}

\begin{figure}
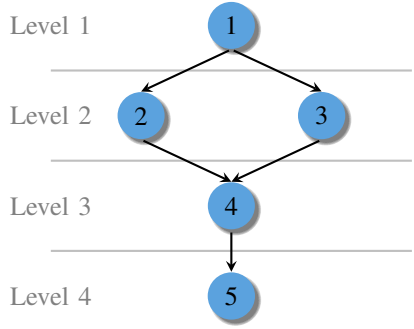

	\centering
	\LUGraphLevels{rwthblue}
	\caption{Result of sorting by levels the graph of dependencies for the \lu{}.} \label{fig:luGraphLevels}
\end{figure}

Once \click{} has generated the dependency graph it
selects all the possible subgraphs that satisfy the dependencies.
Each of the subgraphs corresponds to a different loop-invariant,
provided that it is feasible.
The algorithm starts by sorting the nodes in the dependency
graph; as such a graph is a DAG (direct acyclic graph),
the nodes may be sorted by levels according to the longest path
from the root. For the \lu{} the sorted DAG is shown in 
Fig.~\ref{fig:luGraphLevels}.

\click{} creates the list of subgraphs of the DAG incrementally,
by levels. At first it initializes the
list of subgraphs with the empty subset,
$l = [\{\}]$, which is equivalent to selecting none of 
the PME tasks. Then, at each level it extends the set
of subgraphs by adding all those resulting from appending the accesible nodes
to the existing ones. A node at a given level is accesible from a
subgraph $g$ if all the dependencies of the node are satisfied by $g$.
Fig.~\ref{fig:subDAG} includes a sketch of the algorithm.

In the first iteration of the $LU$ example, 
the only accesible node from $\{\}$ at level 1 is node 1, 
hence, union($\{\}$, $\{1\}$) is added to
$l$ which becomes $[\{\}, \{1\}]$. Now, the
level is increased to 2; no node in level 2 is
accesible from $\{\}$, while both nodes 2 and 3
are accesible from $\{1\}$.
The union of $\{1\}$ with the non-empty subsets of $\{2, 3\}$---$\{2\}$, 
$\{3\}$ and $\{2, 3\}$---are added to $l$, resulting in
$l = [\{\}, \{1\}, \{1, 2\}, \{1, 3\}, \{1, 2, 3\}]$.
At level 3, \click{} discovers that node 4 is accesible from 
subgraph $\{1, 2, 3\}$, thus $\{1, 2, 3, 4\}$ is added
to $l$. Finally, node 5 is accesible from $\{1, 2, 3, 4\}$.
The final list of subgraphs is:
$$[\{\}, \{1\}, \{1, 2\}, \{1, 3\}, \{1, 2, 3\}, \{1, 2, 3, 4\}, \{1, 2, 3, 4, 5\}].$$

The seven subgraphs included in the final list correspond to
predicates that are candidates
to becoming loop-invariants. To this end, \click{} checks each
predicate to establish its feasibility.
The methodology we follow imposes two 
constraints for such a predicate $P$ to be a feasible loop-invariant:
1) there must exist a basic initialization of the operands, 
i.e., an initial partitioning, that renders the predicate $P$ true;
2) $P$ and the negation of the loop-guard, $G$, must imply the 
postcondition, $P_{\rm post}$, of the target operation: 
$P \wedge \neg G \implies P_{\rm post}$.

Following these rules the predicates corresponding to the
empty and the full subgraphs of the DAG are always discarded.
The former because it is analogous to an empty predicate
and no matter what $G$ is, the implication $P \wedge \neg G 
\implies P_{\rm post}$ is not satisfied; the latter 
because it corresponds to the complete computation of the
operation and, therefore, no basic initialization
can be found to render the predicate $P$ true.

\click{} reaches to the same conclusion by identifying
the initial and final state of the partitionings
of the operands and rewriting the predicates in terms of such partitionings.
A detailed discussion through the $LU$ example follows.

\begin{figure}
\begin{center}
\begin{lstlisting}
l = [{}]
for each level i:
  for each subgraph g in l:
    acc = accesiblesNodesFrom(g, i)
    sub = nonEmptySubsets(acc)
    for each s in sub:
      append(l, union(g, s))
  end
end
\end{lstlisting}
\caption{Algorithm to obtain all the possible subgraphs of a DAG.}
\label{fig:subDAG}
\end{center}
\end{figure}

Initially \click{} determines the direction in which
the operands are traversed. In the example, 
all the operands are visited from the top-left
to the bottom-right corner.
The resulting initial partitionings are shown in 
Box~\ref{box:initPart}.

\begin{mybox}
{\small
$$
\renewcommand{\arraystretch}{1.2}
\begin{array}{c}
	\lowtriRuleTwoByTwo{L}{m}{m}{TL}{0}{0} \raisebox{2.8mm}{\textnormal{, }} \hspace{5mm}
	\upptriRuleTwoByTwo{U}{m}{m}{TL}{0}{0} \raisebox{2.4mm}{\textnormal{ and }} \\[10mm]
	\ruleTwoByTwo{A}{m}{m}{TL}{0}{0}
\end{array}
$$
}
\caption{Initial partitioning of the operands for the \lu{}.} \label{box:initPart}
\end{mybox}

\noindent
This knowledge is enough to rule the subgraph $\{1, 2, 3, 4, 5\}$ out;
the application of the rules in Box~\ref{box:initPart} to
the associated predicate $P$

\begin{mybox}
{\small
$$
\renewcommand{\arraystretch}{1.2}
\begin{array}{c}
	\lowtriRuleTwoByTwo{L}{m}{m}{TL}{m}{m} \raisebox{2.8mm}{\textnormal{, }} \hspace{5mm}
	\upptriRuleTwoByTwo{U}{m}{m}{TL}{m}{m} \raisebox{2.4mm}{\textnormal{ and }} \\[10mm]
	\ruleTwoByTwo{A}{m}{m}{TL}{m}{m}
\end{array}
$$
}
\caption{State of the partitioning of the operands for the \lu{}
         upon completion of the loop.} \label{box:finalPart}
\end{mybox}
 
\begin{table*}[!htb] \centering
\begin{tabular}{ccl} \toprule
\raisebox{2mm}{\bf \#} & \multicolumn{1}{c}{\raisebox{2mm}{\bf \footnotesize Subgraph}} & 
\multicolumn{1}{c}{\raisebox{2mm}{\bf \footnotesize Loop-invariant}} \\[-2.5mm]\midrule
1 &
\raisebox{-3.2em}{\smallDepGraphLU{aicesred}{rwthblue}{rwthblue}{rwthblue}{rwthblue}} &
\renewcommand{\arraystretch}{1.4}
%\scriptsize
$
	\left( 
	  \begin{array}{@{\,}c@{\,}|@{\,}c@{\,}}
	  \{ L_{TL}, U_{TL} \} = LU(A_{TL})   & \qquad \, \neq \qquad \phantom{} \\\hline
	     \neq                             & \neq 
	  \end{array} 
	\right)
$ \\
2 &
\raisebox{-3.2em}{\smallDepGraphLU{aicesred}{aicesred}{rwthblue}{rwthblue}{rwthblue}} &
\renewcommand{\arraystretch}{1.4}
%\scriptsize
$
	\left( 
	  \begin{array}{@{\,}c@{\,}|@{\,}c@{\,}}
	  \{ L_{TL}, U_{TL} \} = LU(A_{TL})   & U_{TR} = L_{TL}^{-1} A_{TR} \\\hline 
	     \neq                             & \neq 
	  \end{array} 
	\right)
$ \\
3 &
\raisebox{-3.2em}{\smallDepGraphLU{aicesred}{rwthblue}{aicesred}{rwthblue}{rwthblue}} &
\renewcommand{\arraystretch}{1.4}
%\scriptsize
$
	\left( 
	  \begin{array}{@{\,}c@{\,}|@{\,}c@{\,}}
	  \{ L_{TL}, U_{TL} \} = LU(A_{TL})   & \qquad \, \neq \qquad \phantom{} \\\hline
	     L_{BL} = A_{BL} U_{TL}^{-1}      & \neq 
	  \end{array} 
	\right)
$ \\
4 &
\raisebox{-3.2em}{\smallDepGraphLU{aicesred}{aicesred}{aicesred}{rwthblue}{rwthblue}} &
\renewcommand{\arraystretch}{1.4}
%\scriptsize
$
	\left( 
	  \begin{array}{@{\,}c@{\,}|@{\,}c@{\,}}
	  \{ L_{TL}, U_{TL} \} = LU(A_{TL})   & U_{TR} = L_{TL}^{-1} A_{TR} \\\hline 
	     L_{BL} = A_{BL} U_{TL}^{-1}      & \neq 
	  \end{array} 
	\right)
$ \\
5 &
\raisebox{-3.2em}{\smallDepGraphLU{aicesred}{aicesred}{aicesred}{aicesred}{rwthblue}} &
\renewcommand{\arraystretch}{1.4}
%\scriptsize
$
	\left( 
	  \begin{array}{@{\,}c@{\,}|@{\,}c@{\,}}
	  \{ L_{TL}, U_{TL} \} = LU(A_{TL})   &    U_{TR} = L_{TL}^{-1} A_{TR} \\\hline 
	     L_{BL} = A_{BL} U_{TL}^{-1}      & A_{BR} = A_{BR} - L_{BL} U_{TR} 
	  \end{array} 
	\right)
$ \\\bottomrule
\end{tabular}
\caption{The five loop-invariants for the \lu{}.} \label{tab:LULoopInvs}
\end{table*}

{\scriptsize
\renewcommand{\arraystretch}{1.4}
$$
	\left( 
	  \begin{array}{@{\,}c@{\,}|@{\,}c@{\,}}
	  \{ L_{TL}, U_{TL} \} = LU(A_{TL})   &    U_{TR} = L_{TL}^{-1} A_{TR} \\\hline 
	     L_{BL} = A_{BL} U_{TL}^{-1}      & \{ L_{BR}, U_{BR} \} = LU(A_{BR} - L_{BL} U_{TR}) 
	  \end{array} 
	\right)
$$
}

\noindent
lead to a situation in which all quadrants are empty except for the bottom-right, 
where the computation of the \lu{} of $A_{BR}$ is needed to satisfy $P$.

The initial partitionings determine that the valid loop-guard for the algorithm
is $G = size(A_{TL}) < size(A)$: initially the quadrant $A_{TL}$ is of size 
$0 \times 0$; and at each iteration its size grows until it reaches the same size of $A$. 
The loop-guard $G$ implies that when the loop completes
$A_{TL}$, $L_{TL}$ and $U_{TL}$ are of the same size of $A$, $L$ and $U$.
\click{} exploits this fact to determine the feasibility of a predicate $P$.
It applies the rewrite rules in Box~\ref{box:finalPart}
to $P$ and compares the result to the equation in the postcondition.
Since the result of applying such rules to the empty predicate

{\scriptsize
\renewcommand{\arraystretch}{1.4}
$$
	\left( 
	  \begin{array}{@{\,}c@{\,}|@{\,}c@{\,}}
	    \qquad \, \neq \qquad \phantom{} & \qquad \, \neq \qquad \phantom{} \\\hline
	    \neq                             & \neq 
	  \end{array} 
	\right),
$$
}

\vspace{-3mm}
\noindent
where $\neq$ states that no constraints have to be satisfied,
does not equal the postcondition it is discarded.

The other five predicates satisfy both feasibility constraints and are promoted
to valid loop-invariants for the \lu{}
(Tab.~\ref{tab:LULoopInvs}).
It is important to point out that the five
loop-invariants that \click{} identifies have been well known
for a long time and are commonly presented in linear algebra
textbooks\cite{stew:98a}. At the same time, no explanation relative to
their cardinality is ever provided and, most importantly,
they are presented as distinct entities without a common root.
It is only our systematic methodology that unifies 
these five algorithms for the \lu{}. 

\section{A more complex example: the coupled Sylvester equation}
\label{sec:coupsylv}

As a last study case, we show an example where the complexity
of the graph of dependencies and the number of loop-invariants
are such that the automation becomes an indispensable tool. 
This is by no means the most complex example \click{} may handle,
but a compromise between a relatively complex example and 
the space needed to demonstrate it.
In Box~\ref{box:coupOpDesc}, the \cs{} is defined.

\begin{table*}
\centering
\scriptsize
\renewcommand{\arraystretch}{2.6}
\begin{tabular}{cl} \toprule
\raisebox{2mm}{\bf \#} & \multicolumn{1}{c}{\raisebox{2mm}{\bf \footnotesize Partitioned Matrix Expression}} \\[-2.5mm]\midrule
1 &
$
	\left( 
	  \begin{array}{@{\,}c@{\;}|@{\;}c@{\,}}
	    \{X_{L}, Y_{L} \} = \Psi(A, B_{TL}, C_{L}, D, E_{TL}, F_{L}) & 
	    \{X_{R}, Y_{R} \} = \Psi(A, B_{BR}, C_{R} - Y_{L} B_{TR}, D, E_{BR}, F_{R} - Y_{L} E_{TR})
	  \end{array} 
	\right)
$ \\[5mm]
2 &
$
	\left( 
	  \begin{array}{@{\,}c@{\,}}
	    \{X_{T}, Y_{T} \} = \Psi(A_{TL}, B, C_{T}, D_{TL}, E, F_{T}) \\\hline 
	    \{X_{B}, Y_{B} \} = \Psi(A_{BR}, B, C_{B} - A_{BL} X_{T}, D_{BR}, E, F_{B} - D_{BL} X_{T}) 
	  \end{array} 
	\right)
$ \\[9mm]
3 &
$
	\left( 
	  \begin{array}{@{\,}c@{\;}|@{\;}c@{\,}}
	    \{X_{TL}, Y_{TL} \} = \Psi(A_{TL}, B_{TL}, C_{TL}, D_{TL}, E_{TL}, F_{TL}) & 
		\begin{aligned}
	    \{X_{TR}, Y_{TR} \} = \Psi(& A_{TL}, B_{BR}, C_{TR} - Y_{TL} B_{TR}, \\& D_{TL}, E_{BR}, F_{TR} - Y_{TL} E_{TR})
		\end{aligned} \\\hline 
		\begin{aligned}
	    \{X_{BL}, Y_{BL} \} = \Psi(& A_{BR}, B_{TL}, C_{BL} - A_{BL} X_{TL}, \\& D_{BR}, E_{TL}, F_{BL} - D_{BL} X_{TL}) 
		\end{aligned} &
		\begin{aligned}
	    \{X_{BR}, Y_{BR} \} = \Psi(& A_{BR}, B_{BR}, C_{BR} - A_{BL} X_{TR} - Y_{BL} B_{TR}, \\& D_{BR}, E_{BR}, F_{BR} - D_{BL} X_{TR} - Y_{BL} E_{TR})
		\end{aligned}
	  \end{array}
	\right)
$ \vspace{1mm} \\
\bottomrule
\end{tabular}
\caption{The three Partitioned Matrix Expressions for the \cs{}.} \label{tab:coupsylvPMEs}
\end{table*}

The description in Box~\ref{box:coupOpDesc} is the input 
for \click{}. The system finds three
feasible sets of partitioning rules for the operation. For each of the
sets, \click{} applies the rules to the equation in the postcondition
obtaining a partitioned postcondition. Then, 
the partitioned operands are combined and the equality operator 
is distributed obtaining an expression with multiple equalities.
\click{} takes such expressions and, through a process based on pattern matching 
and algebraic manipulation, obtains the corresponding PMEs.
The three resulting PMEs are listed in Tab.~\ref{tab:coupsylvPMEs}

\begin{mybox}
\footnotesize
$$
\begin{aligned}
\{X, Y\} = \Psi(&A, B, C,\\ &D, E, F)
\end{aligned} \equiv
\left\{
\begin{aligned}
P_{\rm pre}: \{ & \prop{Known}{A} \,\! \wedge \,\! \prop{LowTri}{A} \, \wedge \\
                & \prop{Known}{B} \,\! \wedge \,\! \prop{UppTri}{B} \, \wedge \\
                & \prop{Known}{D} \,\! \wedge \,\! \prop{LowTri}{D} \, \wedge \\
                & \prop{Known}{E} \,\! \wedge \,\! \prop{UppTri}{E} \, \wedge \\
                & \prop{Known}{C} \,\! \wedge %\\
                  \prop{Known}{F} \, \wedge \\
                & \prop{Unknown}{X} \,\! \wedge %\\
                  \prop{Unknown}{Y} \\
\\
P_{\rm post}: & \left\{ \begin{array}{@{}l@{}}
                          A X + Y B = C \\
                          D X + Y E = F 
						\end{array} \right.
\end{aligned}
\right.
$$
\caption{Formal description of the \cs{}.}
\label{box:coupOpDesc}
\end{mybox}

We continue the example by selecting the 
PME in the third row of Tab.~\ref{tab:coupsylvPMEs} and describing the steps performed by \click{}
to obtain loop-invariants. First, the system traverses the PME, one quadrant at a time, to decompose
the equalities into basic tasks. The analysis starts from the top-left equality; since the 
right-hand side consists of a function where all the input arguments are sub-operands,
the system yields the entire expression as a basic task.
\begin{itemize}
\small
\item $\{X_{TL}, Y_{TL} \} := \Psi(A_{TL}, B_{TL}, C_{TL}, D_{TL}, E_{TL}, F_{TL})$.
\end{itemize}

Next, the top-right equality is inspected. In this case, two of the input arguments
are not sub-operands. Thus, \click{} analyzes recursively both arguments,
{\small $C_{TR} - Y_{TL} B_{TR}$} and {\small $F_{TR} - Y_{TL} E_{TR}$},
to identify a sequence of basic tasks. 
The pattern {\small $\tt{A - BC}$}, corresponding
to a basic task, matches both expressions. As a result, \click{} returns
three tasks.
\begin{itemize}
\small
\item $C_{TR} := C_{TR} - Y_{TL} B_{TR}$ \\[-3mm]
\item $F_{TR} := F_{TR} - Y_{TL} E_{TR}$ \\[-3mm]
\item $\{X_{TR}, Y_{TR} \} := \Psi(A_{TL}, B_{BR}, C_{TR}, D_{TL}, E_{BR}, F_{TR})$ \\[-3mm]
\end{itemize}
A similar situation occurs when studying the bottom-left equality,
in which \click{} yields three more basic tasks.
\begin{itemize}
\small
\item $C_{BL} := C_{BL} - A_{BL} X_{TL}$ \\[-2mm]
\item $F_{BL} := F_{BL} - D_{BL} X_{TL}$ \\[-2mm]
\item $\{X_{BL}, Y_{BL} \} := \Psi(A_{BR}, B_{TL}, C_{BL}, D_{BR}, E_{TL}, F_{BL})$  \\[-2mm]
\end{itemize}

Only the equality in the bottom-right quadrant 
remains to be analyzed. \click{} recognizes
that two of the input arguments to the function are not sub-operands.
The difference with the previous two cases is that these two arguments consist
on more than one basic task. For instance, in the expression: {\small 
$C_{BR} - A_{BL} X_{TR} - Y_{BL} B_{TR}$} the pattern {\small $\tt{A - BC}$} matches
{\small $C_{BR} - A_{BL} X_{TR}$} and {\small $C_{BR} - Y_{BL} B_{TR}$}.
\click{} also keeps track of the fact that both tasks are independent from
one another, since they may be computed in any order.
After studying the bottom-right equality, the system yields the following five 
tasks, two per non-basic input argument and the top-level function.
\begin{itemize}
\small
\item $C_{BR} := C_{BR} - A_{BL} X_{TR}$ \\[-3mm]
\item $C_{BR} := C_{BR} - Y_{BL} B_{TR}$ \\[-3mm]
\item $F_{BR} := F_{BR} - D_{BL} X_{TR}$ \\[-3mm]
\item $F_{BR} := F_{BR} - Y_{BL} E_{TR}$ \\[-3mm]
\item $\{X_{BR}, Y_{BR} \} := \Psi(A_{BR}, B_{BR}, C_{BR}, D_{BR}, E_{BR}, F_{BR})$, \\[-3mm]
\end{itemize}
In this last set of returned tasks, 
the first and the second are independent to one another, and
so are the third and the fourth. To summarize, we list the 
twelve basic tasks into which the PME has been decomposed:

\begin{enumerate}
\small
\item $\{X_{TL}, Y_{TL} \} := \Psi(A_{TL}, B_{TL}, C_{TL}, D_{TL}, E_{TL}, F_{TL})$ \\[-3mm]
\item $C_{TR} := C_{TR} - Y_{TL} B_{TR}$ \\[-3mm]
\item $F_{TR} := F_{TR} - Y_{TL} E_{TR}$ \\[-3mm]
\item $\{X_{TR}, Y_{TR} \} := \Psi(A_{TL}, B_{BR}, C_{TR}, D_{TL}, E_{BR}, F_{TR})$ \\[-3mm]
\item $C_{BL} := C_{BL} - A_{BL} X_{TL}$ \\[-3mm]
\item $F_{BL} := F_{BL} - D_{BL} X_{TL}$ \\[-3mm]
\item $\{X_{BL}, Y_{BL} \} := \Psi(A_{BR}, B_{TL}, C_{BL}, D_{BR}, E_{TL}, F_{BL})$  \\[-3mm]
\item $C_{BR} := C_{BR} - A_{BL} X_{TR}$ \\[-3mm]
\item $C_{BR} := C_{BR} - Y_{BL} B_{TR}$ \\[-3mm]
\item $F_{BR} := F_{BR} - D_{BL} X_{TR}$ \\[-3mm]
\item $F_{BR} := F_{BR} - Y_{BL} E_{TR}$ \\[-3mm]
\item $\{X_{BR}, Y_{BR} \} := \Psi(A_{BR}, B_{BR}, C_{BR}, D_{BR}, E_{BR}, F_{BR})$ \\[-3mm]
\end{enumerate}

\begin{mybox*}
\begin{minipage}{0.48\textwidth}
\begin{enumerate}
\small
\item $\{X_{TL}, Y_{TL} \} := \Psi(A_{TL}, B_{TL}, C_{TL}, D_{TL}, E_{TL}, F_{TL})$ \\[-2mm]
\item $C_{TR} := C_{TR} - Y_{TL} B_{TR}$ \\[-2mm]
\item $F_{TR} := F_{TR} - Y_{TL} E_{TR}$ \\[-2mm]
\item $\{X_{TR}, Y_{TR} \} := \Psi(A_{TL}, B_{BR}, C_{TR}, D_{TL}, E_{BR}, F_{TR})$ \\[-2mm]
\item $C_{BL} := C_{BL} - A_{BL} X_{TL}$ \\[-2mm]
\item $F_{BL} := F_{BL} - D_{BL} X_{TL}$ \\[-2mm]
\item $\{X_{BL}, Y_{BL} \} := \Psi(A_{BR}, B_{TL}, C_{BL}, D_{BR}, E_{TL}, F_{BL})$ \\[-2mm]
\item $C_{BR} := C_{BR} - A_{BL} X_{TR}$ \\[-2mm]
\item $C_{BR} := C_{BR} - Y_{BL} B_{TR}$ \\[-2mm]
\item $F_{BR} := F_{BR} - D_{BL} X_{TR}$ \\[-2mm]
\item $F_{BR} := F_{BR} - Y_{BL} E_{TR}$ \\[-2mm]
\item $\{X_{BR}, Y_{BR} \} := \Psi(A_{BR}, B_{BR}, C_{BR}, D_{BR}, E_{BR}, F_{BR})$ \\[-2mm]
\end{enumerate}
\end{minipage}
\hspace{5mm}
\begin{minipage}{0.48\textwidth}
\coupsylvDepGraph{rwthblue}
\end{minipage}
\caption{Graph of dependencies for the \cs{}.} \label{box:coupsylvDepGraph}
\end{mybox*}

Once the equalities are decomposed, \click{} inspects the tasks for dependencies.
Once more, we highlight the dependencies using {\bf boldface}.
The analysis commences from Task 1, whose output sub-operands are $X_{TL}$ and $Y_{TL}$.
$X_{TL}$ is an input for Tasks 5 and 6, while $Y_{TL}$ is an input for Tasks
2 and 3. 
\begin{enumerate}
\small
\item $\{\mathbf{X_{TL}}, \mathbf{Y_{TL}} \} := \Psi(A_{TL}, B_{TL}, C_{TL}, D_{TL}, E_{TL}, F_{TL})$ \\[-3mm]
\item $C_{TR} := C_{TR} - \mathbf{Y_{TL}} B_{TR}$ \\[-3mm]
\item $F_{TR} := F_{TR} - \mathbf{Y_{TL}} E_{TR}$ \\[-3mm]
\setcounter{enumi}{4}
\item $C_{BL} := C_{BL} - A_{BL} \mathbf{X_{TL}}$ \\[-3mm]
\item $F_{BL} := F_{BL} - D_{BL} \mathbf{X_{TL}}$ \\[-3mm]
\end{enumerate}
Therefore, the system identifies a true dependency from Task 1 to each of
Tasks 2, 3, 5 and 6.
Next, \click{} analyzes Task 2, whose output, $C_{TR}$, is an input argument for
Task 4.

\begin{enumerate}
\small
\setcounter{enumi}{1}
\item $\mathbf{C_{TR}} := C_{TR} - Y_{TL} B_{TR}$ \\[-3mm]
\setcounter{enumi}{3}
\item $\{X_{TR}, Y_{TR} \} := \Psi(A_{TL}, B_{BR}, \mathbf{C_{TR}}, D_{TL}, E_{BR}, F_{TR})$ \\[-3mm]
\end{enumerate}
Hence, the corresponding true dependency is imposed.
The analysis continues with Task 3, whose output, $F_{TR}$,
is an input argument of Task 4.

\begin{enumerate}
\small
\setcounter{enumi}{2}
\item $\mathbf{F_{TR}} := F_{TR} - Y_{TL} E_{TR}$ \\[-2mm]
\item $\{X_{TR}, Y_{TR} \} := \Psi(A_{TL}, B_{BR}, C_{TR}, D_{TL}, E_{BR}, \mathbf{F_{TR}})$ \\[-3mm]
\end{enumerate}
As a result, \click{} enforces a true dependency from Task 3 to Task 4. %Also, $Y_{TR}$
The algorithm procedes by analyzing Task 4. One of its output sub-operands, $X_{TR}$,
appears as an input argument of Tasks 8 and 10.

\begin{enumerate}
\small
\setcounter{enumi}{3}
\item $\{\mathbf{X_{TR}}, Y_{TR} \} := \Psi(A_{TL}, B_{BR}, C_{TR}, D_{TL}, E_{BR}, F_{TR})$ \\[-2mm]
\setcounter{enumi}{7}
\item $C_{BR} := C_{BR} - A_{BL} \mathbf{X_{TR}}$ \\[-2mm]
\setcounter{enumi}{9}
\item $F_{BR} := F_{BR} - D_{BL} \mathbf{X_{TR}}$ \\[-3mm]
\end{enumerate}
Two new true dependencies arise: one from Task 4 to Task 8 
and another one from Task 4 to Task 10.

The study of Tasks 5, 6 and 7 is analogous to that of Tasks 2, 3, and 4.
\click{} finds true dependencies from Tasks 5 and 6 to Task 7
\begin{enumerate}
\small
\setcounter{enumi}{4}
\item $\mathbf{C_{BL}} := C_{BL} - A_{BL} X_{TL}$ \\[-2mm]
\item $\mathbf{F_{BL}} := F_{BL} - D_{BL} X_{TL}$ \\[-2mm]
\item $\{X_{BL}, Y_{BL} \} := \Psi(A_{BR}, B_{TL}, \mathbf{C_{BL}}, D_{BR}, E_{TL}, \mathbf{F_{BL}}),$  \\[-3mm]
\end{enumerate}

\noindent
and from Task 7 to Tasks 9 and 11

\begin{enumerate}
\small
\setcounter{enumi}{6}
\item $\{X_{BL}, \mathbf{Y_{BL}} \} := \Psi(A_{BR}, B_{TL}, C_{BL}, D_{BR}, E_{TL}, F_{BL})$  \\[-2mm]
\setcounter{enumi}{8}
\item $C_{BR} := C_{BR} - \mathbf{Y_{BL}} B_{TR}$ \\[-2mm]
\setcounter{enumi}{10}
\item $F_{BR} := F_{BR} - \mathbf{Y_{BL}} E_{TR}.$ \\[-3mm]
\end{enumerate}

\click{} continues the analysis of dependencies with the study of Task
8. Despite that its output, $C_{BR}$, is an input and also the output of
Task 9, there is no dependency between them; during
the decomposition of the corresponding equality, \click{} learned
that they are independent to one another.
Additionally, $C_{BR}$ is also an
input argument for operation 12.

\begin{enumerate}
\small
\setcounter{enumi}{7}
\item $\mathbf{C_{BR}} := C_{BR} - A_{BL} X_{TR}$ \\[-3mm]
\item $C_{BR} := C_{BR} - Y_{BL} B_{TR}$ \\[-3mm]
\setcounter{enumi}{11}
\item $\{X_{BR}, Y_{BR} \} := \Psi(A_{BR}, B_{BR}, \mathbf{C_{BR}}, D_{BR}, E_{BR}, F_{BR})$ \\[-3mm]
\end{enumerate}

\noindent
Consequently, a true dependency is imposed from Task 8 to Task 12.
The very exact same situation is found in the analysis of Task 9.

\begin{enumerate}
\small
\setcounter{enumi}{7}
\item $C_{BR} := C_{BR} - A_{BL} X_{TR}$ \\[-3mm]
\item $\mathbf{C_{BR}} := C_{BR} - Y_{BL} B_{TR}$ \\[-3mm]
\setcounter{enumi}{11}
\item $\{X_{BR}, Y_{BR} \} := \Psi(A_{BR}, B_{BR}, \mathbf{C_{BR}}, D_{BR}, E_{BR}, F_{BR})$ \\[-3mm]
\end{enumerate}

\noindent
A new dependency from Task 9 to Task 12 is established. 
The study of the dependencies for Tasks 10 and 11 is led by the same
principle as for Tasks 8 and 9, originating the corresponding dependencies.
Finally, Task 12 is analyzed. Its output, $\{X_{BR}, Y_{BR} \}$, 
does not appear in any of the other tasks, thus no new dependencies
are imposed. The final graph of dependencies
is shown in Box~\ref{box:coupsylvDepGraph}.

Once the graph is built, \click{} executes the algorithm exposed
in Sec.~\ref{sec:depGraph} returning a list with the predicates
that are canditates to becoming loop-invariants. Then, the predicates are
checked to establish their feasibility; the non-feasible ones
are discarded. In the \cs{} example, the system identifies
64 different loop-invariants, which accordingly will lead to 
64 different algorithms to solve the equation. In Tab.~\ref{tab:coupSylvLinvs}
we list a subset of the returned loop-invariants. 

\begin{table*}[h] \centering
\scriptsize
\begin{tabular}{ccl} \toprule
\raisebox{2mm}{\bf \#} & \multicolumn{1}{c}{\raisebox{2mm}{\bf \footnotesize Subgraph}} & 
\multicolumn{1}{c}{\raisebox{2mm}{\bf \footnotesize Loop-invariant}} \\[-2.5mm]\midrule
1 &
\raisebox{-8.0em}{\smallDepGraphCoupSylv{rwthblue}{rwthblue}{rwthblue}{rwthblue}{rwthblue}} &
\renewcommand{\arraystretch}{2.8}
%\scriptsize
$
	\left( 
	  \begin{array}{@{\,}c@{\;}|@{\;}c@{\,}}
	    \{X_{TL}, Y_{TL} \} = \Psi(A_{TL}, B_{TL}, C_{TL}, D_{TL}, E_{TL}, F_{TL}) & 
	    \phantom{X_{TL}, Y_{TL} } \neq \phantom{X_{TL}, Y_{TL} } \\\hline
		\neq &
		\neq
	  \end{array}
	\right)
$ \\[23mm]
2 &
\raisebox{-8.0em}{\smallDepGraphCoupSylv{aicesred}{rwthblue}{rwthblue}{rwthblue}{rwthblue}} &
\renewcommand{\arraystretch}{2.8}
%\scriptsize
$
	\left( 
	  \begin{array}{@{\,}c@{\;}|@{\;}c@{\,}}
	    \{X_{TL}, Y_{TL} \} = \Psi(A_{TL}, B_{TL}, C_{TL}, D_{TL}, E_{TL}, F_{TL}) & 
	    X_{TR} = C_{TR} - Y_{TL} B_{TR} \\\hline 
		\neq &
		\neq
	  \end{array}
	\right)
$ \\[23mm]
3 &
\raisebox{-8.0em}{\smallDepGraphCoupSylv{rwthblue}{aicesred}{rwthblue}{rwthblue}{rwthblue}} &
\renewcommand{\arraystretch}{2.8}
%\scriptsize
$
	\left( 
	  \begin{array}{@{\,}c@{\;}|@{\;}c@{\,}}
	    \{X_{TL}, Y_{TL} \} = \Psi(A_{TL}, B_{TL}, C_{TL}, D_{TL}, E_{TL}, F_{TL}) & 
	    Y_{TR} = F_{TR} - Y_{TL} E_{TR} \\\hline
		\neq &
		\neq
	  \end{array}
	\right)
$ \\[23mm]
4 &
\raisebox{-8.0em}{\smallDepGraphCoupSylv{rwthblue}{rwthblue}{aicesred}{rwthblue}{rwthblue}} &
\renewcommand{\arraystretch}{2.8}
%\scriptsize
$
	\left( 
	  \begin{array}{@{\,}c@{\;}|@{\;}c@{\,}}
	    \{X_{TL}, Y_{TL} \} = \Psi(A_{TL}, B_{TL}, C_{TL}, D_{TL}, E_{TL}, F_{TL}) & 
	    \phantom{X_{TL}, Y_{TL} } \neq \phantom{X_{TL}, Y_{TL} } \\\hline
	    X_{BL} = C_{BL} - A_{BL} X_{TL} &
		\neq
	  \end{array}
	\right)
$ \\[23mm]
& \LARGE \hspace{-7.7mm} $\vdots$ & \LARGE \hspace{4cm} $\vdots$ \\[4mm]
64 &
\raisebox{-8.0em}{\smallDepGraphCoupSylv{aicesred}{aicesred}{aicesred}{aicesred}{aicesred}} &
\renewcommand{\arraystretch}{2.8}
%\scriptsize
$
	\left( 
	  \begin{array}{@{\,}c@{\;}|@{\;}c@{\,}}
	    \{X_{TL}, Y_{TL} \} = \Psi(A_{TL}, B_{TL}, C_{TL}, D_{TL}, E_{TL}, F_{TL}) & 
		\begin{aligned}
	    \{X_{TR}, Y_{TR} \} = \Psi(& A_{TL}, B_{BR}, C_{TR} - Y_{TL} B_{TR}, \\& D_{TL}, E_{BR}, F_{TR} - Y_{TL} E_{TR})
		\end{aligned} \\\hline 
		\begin{aligned}
	    \{X_{BL}, Y_{BL} \} = \Psi(& A_{BR}, B_{TL}, C_{BL} - A_{BL} X_{TL}, \\& D_{BR}, E_{TL}, F_{BL} - D_{BL} X_{TL}) 
		\end{aligned} &
		\begin{aligned}
	    \{X_{BR}, Y_{BR} \} = \{&C_{BR} - A_{BL} X_{TR} - Y_{BL} B_{TR},\\ &F_{BR} - D_{BL} X_{TR} - Y_{BL} E_{TR}\}
		\end{aligned}
	  \end{array}
	\right)
$ \\\bottomrule
\end{tabular}
\caption{A subset of the 64 loop-invariants for the \cs{}.} \label{tab:coupSylvLinvs}
\end{table*}

The large number of identified loop-invariants and
the corresponding algorithms, demonstrates the necessity 
for having a system that
automates the %entire 
process. As Gries and Schneider point out in his book
{\em A Logical Approach to Discrete Math}~\cite{GrSc:92}

\vspace{2mm}
{\em ``Finding a suitable loop-invariant is the most difficult part of writing
most loops.''} %~\cite{GrSc:92}.}

\section{Conclusions} \label{sec:conclusions}

The results we presented in this paper, in conjunction
with our previous work on PME generation~\cite{CASC-2011-PME}, 
constitute a tangible
step forward towards the automatic generation of
algorithms and code for matrix equations.
We have shown how \click{}, the symbolic system
we developed, identifies loop-invariants for
a target equation from its PMEs  
through a sequence of steps involving pattern
matching and rewrite rules. It is thanks
to a computer algebra system like Mathematica that
such steps are performed automatically.

In order to obtain loop-invariants, \click{}
first breaks down the operations specified in the PME into
a list of basic computational tasks. 
To this end, \click{} analyzes the structure of the
expressions that appear in the PMEs;
this step involves an extensive usage of pattern matching.
In a second step, the resulting tasks are then inspected and a
graph of dependencies is built. Both these
steps heavily rely on the
pattern matching capabilities of Mathematica.
Finally, the system traverses the dependency graph,
selecting the feasible loop-invariants.

We believe the approach to be fairly general, as the
examples provided suggest: even though the \lu{} and the \cs{}
differ in number of operands, complexity and computation;
the steps towards the loop-invariants are exactly the same.
When applied to the \lu{}, \click{} discovers all the known
algorithms and unifies them under a common root. For the \cs{}
instead, \click{} goes well beyond the known algorithms discovering
dozens of new ones.

\section{Acknowledgements}
\sloppypar
The authors wish to thank Matthias Petschow and Roman Iakymchuk
for discussions. Financial support from the
Deutsche Forschungsgemeinschaft (German Research Association) through
grant GSC 111 is gratefully acknowledged.

%\bibliographystyle{IEEEtran}
%\bibliography{biblio}

\end{document}

%% file: rules.tex
\newcommand{\upptriRuleTwoByTwo}[6]
{
  \begin{aligned}
    \renewcommand{\arraystretch}{1.2}
      #1_{#2 \times #3} 
	  \rightarrow &
	  \left( 
	    \begin{array}{@{}c@{\,}|@{\,}c@{}} 
		  #1_{TL} & #1_{TR} \\\hline 
		  0       & #1_{BR} 
		\end{array} 
	  \right) \\
      \small \textnormal{where } & #1_{#4} \textnormal{ is } #5 \times #6
  \end{aligned}
}

\newcommand{\lowtriRuleTwoByTwo}[6]
{
  \begin{aligned}
    \renewcommand{\arraystretch}{1.2}
      #1_{#2 \times #3} 
	  \rightarrow &
	  \left( 
	    \begin{array}{@{}c@{\,}|@{\,}c@{}} 
		  #1_{TL} & 0 \\\hline 
		  #1_{BL} & #1_{BR} 
		\end{array} 
	  \right) \\
      \small \textnormal{where } & #1_{#4} \textnormal{ is } #5 \times #6
  \end{aligned}
}

\newcommand{\ruleTwoByTwo}[6]
{
  \begin{aligned}
    \renewcommand{\arraystretch}{1.2}
      #1_{#2 \times #3} 
	  \rightarrow &
	  \left( 
	    \begin{array}{@{}c@{\,}|@{\,}c@{}} 
		  #1_{TL} & #1_{TR} \\\hline 
		  #1_{BL} & #1_{BR} 
		\end{array} 
	  \right) \\
      \small \textnormal{where } & #1_{#4} \textnormal{ is } #5 \times #6
  \end{aligned}
}

\newcommand{\ruleTwoByOne}[6]
{
  \begin{aligned}
    \renewcommand{\arraystretch}{1.2}
      #1_{#2 \times #3} 
	  \rightarrow &
	  \left( 
	    \begin{array}{@{}c@{}} 
		  #1_{T} \\\hline 
		  #1_{B}
		\end{array} 
	  \right) \\
      \small \textnormal{where } & #1_{#4} \textnormal{ is } #5 \times #6
  \end{aligned}
}

\newcommand{\ruleOneByTwo}[6]
{
  \begin{aligned}
    \renewcommand{\arraystretch}{1.2}
      #1_{#2 \times #3} 
	  \rightarrow &
	  \left( 
	    \begin{array}{@{}c@{\,}|@{\,}c@{}} 
		  #1_{L} & #1_{R}
		\end{array} 
	  \right) \\
      \small \textnormal{where } & #1_{#4} \textnormal{ is } #5 \times #6
  \end{aligned}
}

\newcommand{\ruleOneByOne}[3]
{
  \begin{aligned}
    \renewcommand{\arraystretch}{1.2}
      #1_{#2 \times #3} 
	  \rightarrow &
	  \left( 
	    \begin{array}{@{}c@{}} 
		  #1
		\end{array} 
	  \right) \\
      \small \textnormal{where } & #1 \textnormal{ is } #2 \times #3
  \end{aligned}
}

%% file: inputs/skeleton.tex
\newcommand{\PBefore}{ P_{\rm before} }
\newcommand{\PAfter} { P_{\rm after}  }
\newcommand{\PPost}  { P_{\rm post}   }
\newcommand{\PPre}   { P_{\rm pre}    }
\newcommand{\PInv}   { P_{\rm inv}    }
\newcommand{\EInv}   { E_{\rm inv}    }
\newcommand{\EBefore}{ E_{\rm before} }
\newcommand{\EAfter} { E_{\rm after}  }

\newcommand{\WSupdate} { \rm Loop Body }

\begin{mybox}%[tb]
\centering
  %\begin{center}
    %\renewcommand{\WSrepartition}{\vspace*{-3mm}}
    %\renewcommand{\WSmoveboundary}{\vspace*{-1mm}}
    %\vspace*{-3mm}
    %\vspace*{-1mm}
    % \renewcommand{\WSrepartitionsizes}{$\dots$}
    \begin{tabular}{ l }
    %\begin{tabular}{| @{\hspace*{3.5cm}} l @{\hspace*{3.2cm}} |} \hline
      \\[-2mm]
      $\{ \PPre \}$ \\[1.5mm]
      {\bf Partition} \\[1.5mm]
      $\{ \PInv \}$ \\[1.5mm]
      {\bf While} $G$ {\bf do}\\[1.5mm]
      %\hspace*{8mm}{\bf Repartition} \ $\dots$ \ ; \\[1.5mm]
      \hspace*{7.5mm}$\WSupdate$ \\[1.5mm]
      %\hspace*{8mm}{\bf Continue} \ $\dots$ \ ;\\[1.5mm]
      {\bf end} \\[1.5mm]
      $\{ \PInv \wedge \neg G \}$ \\[2mm]
      $\{ \PPost \}$ \\[1.5mm]
      %\hline
    \end{tabular}
  %\end{center}
  \caption{Template for a formal proof of correctness for algorithms
    consisting of an initialization step followed by a loop.}
  \label{box:skeleton}
\end{mybox}